\documentclass[twocolumn,,showpacs,nofootinbib,prd,aps,amsmath,amssymb,superscriptaddress]{revtex4-1}
\usepackage{bm}% bold math

\usepackage{graphicx,epsfig}% Include figure files
\usepackage{color}

\usepackage[para,online,flushleft]{threeparttable}
\usepackage{multirow}

\newcommand{\ba}{\begin{eqnarray}}
\newcommand{\ea}{\end{eqnarray}}

\newcommand{\B}{{\cal{B}}}

\newcommand{\DD}{{\cal {D}}}

\newcommand{\bbq}{\begin{quote}}
\newcommand{\eeq}{\end{quote}}

\newcommand{\tbb}{t_{\textrm{\tiny{bb}}}}

\newcommand{\tcoll}{t_{\textrm{\tiny{coll}}}}
\newcommand{\tmax}{t_{\textrm{\tiny{max}}}}

\newcommand{\RR}{{}^{(3)}{\cal{R}}}

\newcommand{\FF}{{\cal{F}}}
\newcommand{\GG}{{\cal{G}}}

\newcommand{\VV}{{\cal{V}}}
\newcommand{\HH}{{\cal{H}}}
\newcommand{\KK}{{\cal{K}}}
\newcommand{\PP}{{\cal{P}}}
\newcommand{\QQ}{{\cal{Q}}}

\newcommand{\taubb}{\tau_{\textrm{\tiny{bb}}}}
\newcommand{\taucoll}{\tau_{\textrm{\tiny{coll}}}}
\newcommand{\taumax}{\tau_{\textrm{\tiny{max}}}}
\newcommand{\dkappa}{\delta^{(\kappa)}}
\newcommand{\dmu}{\delta^{(\mu)}}
\newcommand{\Drho}{\Delta^{(\rho)}}
\newcommand{\DKK}{\Delta^{(\KK)}}
\newcommand{\dKK}{\delta^{(\KK)}}
\newcommand{\drho}{\delta^{(\rho)}}
\newcommand{\dbeta}{\delta^{(\beta)}_0}
\newcommand{\DDa}{{\textrm{\bf{D}}}^{(A)}}
\newcommand{\DDrho}{{\textrm{\bf{D}}}^{(\rho)}}
\newcommand{\DDKK}{{\textrm{\bf{D}}}^{(\KK)}}
\newcommand{\DDH}{{\textrm{\bf{D}}}^{(H)}}

\newcommand{\bW}{{\rm{\bf W}}}

\newcommand{\dd}{{\rm{d}}}

\begin{document}
\title[Black hole formation from the gravitational collapse of a non--spherical network of structures]{Black hole formation from the gravitational collapse of a non--spherical network of structures}
\author{Ismael Delgado Gaspar}
\email{ismael.delgado@correo.nucleares.unam.mx}
\affiliation{Instituto de Ciencias Nucleares, Universidad Nacional Aut\'onoma de M\'exico (ICN-UNAM),
A. P. 70--543, 04510 M\'exico D. F., M\'exico.}
\author{Juan Carlos Hidalgo}
\email{hidalgo@fis.unam.mx}
\affiliation{Instituto de Ciencias F\'\i sicas, Universidad Nacional Aut\'onoma de M\'exico, 62210 Cuernavaca, Morelos, M\'exico,}
\author{Roberto A. Sussman}
\email{sussman@nucleares.unam.mx}
\affiliation{Instituto de Ciencias Nucleares, Universidad Nacional Aut\'onoma de M\'exico (ICN-UNAM),
A. P. 70--543, 04510 M\'exico D. F., M\'exico.}
\author{ Israel Quiros}
\email{iquiros@fisica.ugto.mx}
\affiliation{Dpto. Ingenier\'ia Civil, Divisi\'on de Ingenier\'ia, Universidad de Guanajuato, Gto., M\'exico.}
\date{\today}
\begin{abstract}
We examine the gravitational collapse and black hole formation of multiple non--spherical configurations constructed from Szekeres dust 
models with positive spatial curvature that smoothly match to a Schwarzschild exterior.  These configurations are made of an almost 
spherical central core region surrounded by a network of ``pancake--like'' overdensities and voids with spatial positions 
prescribed through standard initial conditions. We show that a full collapse into a focusing singularity, without shell crossings appearing 
before the formation of an apparent horizon, is not possible unless the full configuration becomes exactly or almost spherical.  
Seeking for black hole formation, we demand that shell crossings are covered by the apparent horizon. This
requires very special fine--tuned initial conditions that impose very strong and 
unrealistic constraints on the total black hole mass 
and full collapse time. As a consequence,  non--spherical non--rotating dust sources cannot furnish even minimally realistic toy models of 
black hole formation at astrophysical scales:  demanding realistic collapse time scales yields huge unrealistic black hole masses, 
while simulations of typical astrophysical black hole masses collapse in unrealistically small times. We note, however, that the resulting time--mass constraint is 
compatible with early Universe models of primordial black hole formation, suitable in early dust--like environments. Finally, we argue
 that the shell crossings appearing when non--spherical dust structures collapse are an indicator that such structures do not form galactic mass black holes but virialise into stable stationary objects.   
\end{abstract}
\pacs{04.20.Jb,04.20.-q, 95.35.+d, 97.60.Lf}
%04.20.Jb Exact solutions
%04.20.-q	Classical general relativity
%95.35.+d Dark matter
%97.60.Lf Black holes
%98.80.?k Cosmology
%
\maketitle
\section{Introduction}
The gravitational collapse and black hole (BH) formation (including singularity censorship issues) of spherically symmetric dust models 
has been extensively examined  \cite{datt1938klasse,*oppenheimer1939continued,*waugh1988strengths,*joshi1993naked,*joshi2002naked,*joshi2015all,*eardley1979time,*christodoulou1984violation,*newman1986strengths,*joshi1992structure,*dwivedi1994occurrence,*magli1997gravitational,*magli1998gravitational,*harada1998final,*harada1999nakedness,*da2000collapsing,*harada2001convergence,*goswami2002role,*goncalves2002spectrum,*giambo2003new,*giambo2004naked,*goswami2004gravitational,*satin2016genericity}.
%(\textbf{dust sources, perfect fluid sources}) 
These models are described by the Lema\^\i tre--Tolman--Bondi (LTB) solutions and typically consider a dust over--density (a local spatial density maximum) around the symmetry centre. 

However, the proper study of BH formation from the collapse of non--spherical dust configurations remains an open problem 
%reviews
(see \cite{joshi2011recent,gundlach2007critical}). In particular, the quasi--spherical Szekeres solutions of class I \cite{kras1,kras2,BKHC2009}
 %Kras' books
 allow for modelling non--trivial non--spherical configurations, involving a spheroidal over--density or density void surrounded by 	
 elaborated networks of ``pancake--like structures''. Here, by ``pancake--like structures'' we mean elongated regions that contain local 
 spatial density maxima (over--densities) or minima (voids), which can be localised in terms of radial and angular coordinates of suitable 
 spherical comoving coordinates. As shown in \cite{Bolejko:2006bb,bolejko2007evolution,BoSu2011,multi}, 
% \textbf{citar 26 y 27 de multi en Arxiv}, 
 these Szekeres models allow for prescribing the spatial location of all these extremes from specified initial conditions.

Since quasi--spherical Szekeres models are the least idealised exact solution applicable to cosmology, there is a large body of literature employing them as toy models for structure formation and for fitting cosmological 
observations 
\cite{Bolejko:2006bb,bolejko2007evolution,Hellaby:2007hq,BoSu2011,multi,WH2012,Krasinski:2008jp,*Ishak:2007rp,*Bolejko:2008xh,*Bolejko:2010eb,*Krasinski:2010rc,*Nwankwo:2010mx,*Ishak:2011hz,*Peel:2012vg,*Ishak:2013vha,*Koksbang:2015ima,*Koksbang:2015jba,*Bolejko:2015gmk,*Krasinski:2016jzk,*Krasinski:2017uht,*Hellaby:2017soj,*Bolejko:2017lai}.
%\cite{Bolejko:2006bb,bolejko2007evolution,Hellaby:2007hq,*Krasinski:2008jp,*Ishak:2007rp,*Bolejko:2008xh,*Bolejko:2010eb,*Krasinski:2010rc,*Nwankwo:2010mx,*Ishak:2011hz,*Peel:2012vg,WH2012,*Ishak:2013vha,*Koksbang:2015ima,*Koksbang:2015jba,*Bolejko:2015gmk,paperJCAP,*Krasinski:2016jzk,*Krasinski:2017uht,*Bolejko:2017lai}. 
However, the proper study of BH 
formation from quasi--spherical Szekeres models, and indeed from any non--spherical progenitors, remains largely unexplored. 
In this context, reference \cite{Harada:2015ewt} discusses 
the conditions for BH formation from the collapse of Szekeres configurations, while the definition of their apparent horizon is discussed in \cite{Krasinski:2012hv}. 
These references stand as valuable precedents, but still leave important issues to be examined. 
In particular, in astrophysical systems it is plausible to match the Szekeres central solution to a Schwarzschild exterior. Therefore from the outside the process is seen as the usual spherical collapse. However, we are interested in the non--spherical interior and the evolution of multiple (pancake--like and spherical) structures. The aim of the present article is to explore the collapse of networks of non--spherical  
structures modelled by Szekeres solutions into a single ``Big Crunch'' singularity (final focusing singularity).

We find advantageous to address the problem employing quasi--local scalar variables  adapted to Szekeres models 
 (a formalism developed in \cite{sussbol,multi,paperJCAP,CPTSzek2017}).
 Such formalism is idoneous to describe the complex radial and angular dependence of the density associated with these networks of structures, and their specification through initial conditions. Besides these advantages, the q--scalars and their fluctuations are exact generalisations of cosmological dust perturbations in the synchronous (and comoving) gauge of cosmological perturbation theory 
 \cite{perts,CPTSzek2017}. 

In previous work \cite{paperJCAP} we were concerned with cosmic structure modelling, looking at localised collapsing regions within models whose cosmic background (a $\Lambda$CDM background) is expanding. Consequently we 
considered only two types of ``collapse morphologies'' (defined by the three eigenvalues of the expansion tensor): the ``spherical'' collapse (all negative eigenvalues) and the ``pancake'' collapse (two  positive and one negative eigenvalues). 
Under this approach we simply assumed 
that locally collapsing regions (spherical or pancake) would virialise into stationary stable structures and thus ignored their terminal 
evolution into singularities (Big Crunch or shell crossings).  

In this paper we are interested in astrophysical BH formation from multiple overdensities. We model local collapse (with $\Lambda=0$) of configurations with positive spatial curvature consisting of a central LTB inhomogeneity, surrounding Szekeres pancake solutions and embedded in an exterior Schwarzschild spacetime (see Fig.~\ref{DC3d} below).  We simulate the gravitational collapse through examples evolving the pure growing-mode of the Szekeres structures and find that a full ``Big Crunch'' collapse without shell crossings appearing before the formation of an apparent horizon is not possible unless the full configuration becomes exactly or almost spherical. This is a consequence of the fact that conditions for avoiding shell crossings are much more stringent in Szekeres models than in LTB models. Our results indicate that the setup may represent a suitable model for large--scale structure formation in which the dust structures eventually enter a stage of virialisation beyond the Szekeres description \cite{PadmanabhanBook,*Angrick:2010qg}.

Looking for the possibility of BH formation of fully non--spherical configurations with this proviso, we fine--tune the initial conditions, 
so that shell crossings become covered by the apparent horizon and lie very close to (what would be) the locus of the Big Crunch. 
For such examples, we compute the final collapse time and total BH mass. Our results show either a very short time of collapse 
or a very large mass of the BH developed over astrophysical timescales. 
Instead, our results show compatibility with the theory of primordial black holes (PBH) formation, which involves a rapid collapse 
of very small masses  \cite{Musco:2004ak,*Musco:2008hv,Harada:2015ewt,Harada:2016mhb}.  
%Musco:2008hv,Musco:2004ak

The plan of the paper is as follows. In Section \ref{Sec:SphCoord} we introduce a description of the Szekeres models in terms of q--scalars and spherical coordinates and comment on sufficient conditions for the existence of multiple spatial extrema of the Szekeres scalars. 
General features of the quasi--spherical Szekeres models 
are reviewed in Section \ref{Sec:GravCollVirialization} including the collapse morphologies, a criterium for the identification 
of apparent horizons, and the occurrence of shell crossings and concavity inversions points 
(the evolving of local density  maxima into local minima and vice versa).   
In this Section we argue that shell crossings are indicative of the start of virialisation, therefore we can model the structure formation
process. 
To illustrate our setup, in Section \ref{Sec:SettingUp} we show two representative examples of structure formation with Szekeres models, namely, a galaxy supercluster and a BH. 
Our results are summarised and discussed in Section \ref{Sec:DiscFinalRemarks}. 
Finally, we have included four appendices that complement the main text. Appendices \ref{SysOfDiffEq} and \ref{AnalyticSols}
 provide the evolution equations of the q--scalars and metric functions in Szekeres models and their exact solution for $\Lambda=0$, 
 respectively. These solutions are re--written in a dimensionless form in Appendix \ref{App:C}, and in Appendix \ref{AvoidanceofShx}
we list the general conditions to avoid shell crossings.

\section{Szekeres models in spherical coordinates}\label{Sec:SphCoord}

The  quasi--spherical Szekeres models of class I\footnote{
All further mention of ``Szekeres models'' will refer only to quasi--spherical models of class I (see \cite{kras2} for 
a broad discussion on their classification). 
%Metric determinant and inverse are given in the Appendix. 
We are not considering models whose constant time slices have spherical or wormhole topology \cite{kras2} 
(the appropriate  form of the metric (\ref{szmetric}) for those cases is given in Appendix D of \cite{sussbol}).
}
in terms of ``stereographic'' spherical coordinates are described by the metric \cite{kras2}, 
\ba \dd s^2 -\dd t^2 +a^2\,h_{ij}\,\dd x^i\,\dd x^j,\qquad i,j=r,\theta,\phi,\label{szmetric}\ea
where $a=a(t,r)$ and 
%
%\bse
\ba
 h_{rr} &=& \frac{(\Gamma-\bW)^2}{1-\KK_{qi}r^2}+(\PP+\bW_{,\theta})^2+U^2\bW_{,\phi}^2,\label{szmetric1}
 \\
h_{r\theta} &=& - r\,(\PP+\bW_{,\theta}),\label{szmetric2}
 \\
h_{r\phi} &=&  -r\,U\,\bW_{,\phi},\quad h_{\theta\theta} = r^2,\quad h_{\phi\phi}=r^2\,\sin^2\theta,\label{szmetric3}
\ea
%\ese
%
with
\ba 
\Gamma &=& 1+\frac{ra'}{a},\qquad\,\, U=1-\cos\theta,\label{aux1}
\\
\PP& = &X\cos\phi+Y\sin\phi,\quad \bW=-\PP\sin\theta-Z\cos\theta,\label{aux2}
\ea
and four free parameters $X,\,Y,\,Z,\,\KK_{qi}$ which depend only on $r$ (see interpretation of $\KK_{qi}$ in (\ref{qscals})).  The function $\bW$ has the mathematical structure of a dipole and governs the deviation from spherical and axial symmetries \cite{multi}. Therefore, different particular cases follow by specialising this function:  $X=Y=Z=\bW=0$ corresponds to the spherically symmetric LTB models, while $X=Y=0$, $Z\ne 0$ so that $\bW=\bW(r,\theta)=-Z\cos\theta$ corresponds to axial symmetry. 

\subsection{Quasi--local scalars and their fluctuations}\label{qscalarFlucDef}

To look at the dynamics of the models we introduce the quasi--local variables (q--scalars) $A_q$ for each covariant scalar $A=\rho,\,H=\Theta/3,\,\KK=\RR/6,$ (density, Hubble expansion and spatial curvature) 
\ba 
A_q &=& \frac{\int_\DD{A\,F\,\dd \VV_p}}{\int_\DD{F\,\dd \VV_p}},  \quad \hbox{with} \label{Aqdef}
\\
\dd\VV_p &=& \sqrt{\hbox{det}(g_{ij})}\,\dd^3x=\frac{a^3\,r^2\,(\Gamma-\bW)\,\sin\theta}{\sqrt{1-\KK_{qi}r^2}}\,\dd r\dd\theta\dd\phi,
\nonumber
 \ea
while their exact fluctuations ($\DDa$) are given by \cite{sussbol},
\ba 
 \DDa &=& A-A_q=\frac{r\,A'_q}{3(\Gamma-\bW)},\label{DaDrho}
 \\
  \Drho &=& \frac{\DDrho}{\rho_q}=\frac{\rho-\rho_q}{\rho_q},\label{DaDrho2}
 \ea
which lead to the following scaling laws\footnote{The integral in (\ref{Aqdef}) is evaluated in an arbitrary time slice (constant $t$) in a spherical comoving domain $\DD$ bounded by an arbitrary fixed $r>0$. The lower bound is the locus $r=0$, analogous to the symmetry centre of spherical models \cite{BoSu2011}. While Szekeres models are not spherically symmetric, the surfaces of constant $r$ are non--concentric 2--spheres \cite{kras2}. Notice that $A_q=A_q(t,r)$ even if the scalars $A$ depend on the four coordinates $(t,r,\theta,\phi)$ \cite{sussbol}. 
%Their relation with the average integrals is discussed in section \ref{contrast}.
}:
\ba 
\rho_q &=& \frac{\rho_{qi}}{a^3},\quad \KK_q=\frac{\KK_{qi}}{a^2},\quad H_q=\frac{\dot a}{a},\label{qscals}
\\
 1+\Drho &=& \frac{1+\Drho_i}{\GG},\quad \frac{2}{3}+\DKK =\frac{\frac{2}{3}+\DKK_i}{\GG}, \label{qperts,perts1}
 \\
 \quad \GG &=& \frac{\Gamma-\bW}{1-\bW}.\label{qperts,perts2}
\ea      
Here we have assumed the radial coordinate gauge $a_i=\Gamma_i=\GG_i=1$ with ${}_i$ denoting evaluation at an arbitrary $t=t_i$.
%\textbf{JCH: agregar relaci\'on a perturbaciones con modelos Szekeres}. 

The q--scalars and their fluctuations are covariant objects \cite{sussbol} reduced in the linear limit 
to standard variables of cosmological dust perturbations in the synchronous
gauge \cite{perts,CPTSzek2017}.

\subsection{Spatial location of the  extrema of the Szekeres scalars}

The spatial location of the scalars extrema follows from the condition $A^{\prime}=A_{,\theta}=A_{,\phi}=0$, whose solutions are,
\begin{equation}
r=r_{e \pm}, \; \theta_{\pm}( r_{e \pm}), \; \phi_{\pm}( r_{e \pm}), \;\; \hbox{at $t$=const.}
\end{equation}
where the angular extrema are given by,
\ba  
%\fl 
 \phi_{-} &=& \arctan \left(\frac{Y}{X}\right),\quad \phi_{+} = \pi+\phi_{-}, \label{sol1}\\
 %\fl 
\theta_{-}&=& \arccos\left(\frac{Z}{\sqrt{X^2+Y^2+Z^2}}\right), \quad \theta_{+}= \pi-\theta_{-}.\label{sol2}
 \ea
The extrema define an angular direction for every fixed $r$, and the two ``curves of angular extrema'' $\B_{\pm}(r )=[r,\,\theta_{\pm}( r),\,\phi_{\pm}( r)],$ 
parametrised by $r$ in all time slices. 
 
A sufficient condition for the existence of an arbitrary number of radial extrema of the Szekeres scalars is achieved by assuming a
sequence of ``local homogeneity spheres'', defined by the vanishing at all times of the shear and electric Weyl tensors along a comoving 2--sphere 
of generic radius $r_*$ \cite{multi}. Since the Local Homogeneity Spheres are preserved by the time evolution, they can be specified by initial 
conditions such that 
all the exact fluctuations vanish at $r_*$: $\DDa(t,r_*)=0 \Rightarrow  A'_{q}(t,r_*)=0$.

In general, sufficient conditions for the existence of extrema can be summarised as follows \cite{multi}
\begin{itemize}
\item If regularity conditions hold, the origin of coordinates is a spatial extremum of the scalars. It will be a minimum (void) if 
$A''_{q}(t_0, r=0)>0$ or a maximum (overdensity) if $A''_{q}(t_0, r=0)<0$.
\item There is a radial extremum of the scalars in the radial interval between two homogeneity spheres: $\Delta_*^i=r_*^{i-1}<r<r_*^i$. 
This is a maximum or a minimum depending on the sign of $A'_{q}(t_0,r)$ in  $\Delta_*^i$. 
\item The angular extrema of the scalars lie in the branch $\B_{+}(r)$ of the curves of angular extrema,  
while the other branch only contains saddle points.
\end{itemize}

These  extrema  are  preserved  throughout time evolution, pending shell crossings or concavity inversions which we discuss in Sec. 
\ref{Sec:GravCollVirialization}.

\subsection{The dynamics of the models}   

The dynamics of the models can be fully determined by solving the evolution equations for the variables \eqref{Aqdef}--\eqref{DaDrho2} (see Appendices \ref{SysOfDiffEq} and \ref{AnalyticSols}). 
However, we can also determine the models through their metric functions, in particular the metric function $a$ (which generalises the FLRW scale factor) 
follows from solving the Friedman equation that results from (\ref{constraints1}) and (\ref{qscals}):
\ba t-\tbb =  \int_0^a{\frac{\sqrt{\xi}\,\dd\xi}{\left[\frac{8\pi}{3} \rho_{qi}-\KK_{qi}\,\xi+\frac{8\pi}{3} \Lambda\,\xi^3\right]^{1/2}}},\label{quadrature1}\ea
where $\tbb=\tbb(r)$ is the inhomogeneous ``Big Bang time'' satisfying $a(\tbb(r),r)=0$, which can be found from evaluating the integral (\ref{quadrature1}) for $t=t_i$ up to $a_i=1$. The other scale factor $\Gamma$ in (\ref{aux1}) follows by differentiating both sides of (\ref{quadrature1}) 
and rearranging terms \cite{sussbol}. Once $a$ and $\Gamma$ are found from the quadrature (\ref{quadrature1}) we have analytic expressions for all relevant variables. 
If the cosmological constant is neglected, the quadrature (\ref{quadrature1}) is expressible in terms of elementary functions, 
which leads to analytic solutions of the evolution equations in terms of the scale factors and scaling
laws for the q--scalars and their fluctuations. These solutions are given in detail in Appendix \ref{AnalyticSols}. 

Since Szekeres dust models are characterised by all vorticity, 4--acceleration and magnetic Weyl tensor vanishing together, 
they belong to a class of  
models called ``silent universes'' \cite{SilentUnivPantano,SilentUnivvanElst}, in which no information is propagated either by sound or gravitational 
waves and, consequently, each worldline evolves independently.
The quasi--spherical Szekeres spacetime can be matched either to an FLRW or  
(de Sitter--)Schwarzschild spacetime \cite{kras2,Hellaby:2007hq} and 
due to its silent properties the Szekeres evolution is not affected 
by our background choice.

\section{Gravitational collapse and virialisation}\label{Sec:GravCollVirialization}

\subsection{Singularities and collapse morphologies} 

The collapse morphologies can be described 
%in a coordinate independent manner 
through the ``expansion tensor'' $\HH_{ab}=h_a^c h_b^d u_{c;d}=H\,h_{ab}+\sigma_{ab}$, where $h_{a b}$ and $\sigma_{ab}$
are the spatial projection and shear tensors, respectively.
The tensor $\HH^a_b$ admits three eigenvalues:
\begin{equation}\hbox{\bf{H}}_{(1)}=\hbox{\bf{H}}_{||}=\frac{\dot a}{a}+\frac{\dot\GG}{\GG},\qquad \hbox{\bf{H}}_{(2)}=\hbox{\bf{H}}_{(3)}=\hbox{\bf{H}}_{\perp}=\frac{\dot a}{a},\label{eigenvals}\end{equation} 
which follow from expressing this tensor in terms of a canonical orthonormal triad of spacelike unit vectors \cite{SilentUnivPantano}.
Notice that $H=(\HH^a_a)=\hbox{\bf{H}}_{||}+2\hbox{\bf{H}}_{\perp}$ is a measure of the average expansion/collapse rate.  
These eigenvalues define three normalised ``scale factors'' $\{\ell_{(1)},\,\ell_{(2)},\,\ell_{(3)}\}$ fulfilling 
$\hbox{\bf{H}}_{||}=\dot\ell_{||}/\ell_{||},\,\,\hbox{\bf{H}}_{\perp}=\dot\ell_{\perp}/\ell_{\perp}$: 
\ba
 \ell_{||}=\ell_{(1)} &=&a\GG=\frac{a(\Gamma-\bW)}{1-\bW}, \label{eq:eigenvalues1}
 \\
\ell_{\perp}&=&\ell_{(2)} =\ell_{(3)} =a, \label{eq:eigenvalues2}
\ea 
that describe the rate of local expansion/collapse of dust elements along the principal directions, leading to the following collapse morphologies:
\begin{description}
\item[Spherical collapse] 3--dimensional collapse in which the three scale factors decrease at a similar rate:  $\ell_{(1)},\, \ell_{(2)},\,\ell_{(3)}\to 0$. For these conditions to occur simultaneously we require $a(r,t_{\rm col})\to 0$ at the point of collapse.
%, as can be read in Eq.~\eqref{eq:eigenvalues1}--\eqref{eq:eigenvalues2}. 	
%
\item[``Pancake'' collapse] 1--dimensional collapse 
%along the direction $\hat\eb_{(1)}^a$, 
with $\ell_{||}\to 0$ or decreasing close 
to zero and $\hbox{\bf{H}}_{||}$ becoming very small or negative, with $\ell_{\perp}\gg \ell_{||}$. 
\item[Filamentary collapse] collapse along two principal directions, 
%$\hat\eb_{(2)}^a,\,\hat\eb_{(3)}^a$, 
hence: $\ell_{\perp}\to 0$ with finite (or diverging) $\ell_{||}$.   
\end{description}

\subsection{Shell crossings singularities and concavity inversions} 
\subsubsection{Shell crossings singularities} \label{Sec:Shx}
%\newline

These singularities occur when the mass density diverges as the proper distance between comoving layers 
(with different comoving coordinates $r$) vanishes, while their area distances $R=a\,r$ remain greater than zero.
%In general, they are marked by coordinate surfaces that do not coincide with specific comoving layers and for which the spherically 
%symmetric ``area distance'' $R=a\,r$ is nonzero. 
Shell crossings are considered weak singularities or less severe than 
the Big Bang or Big Crunch (which occur as $a\to 0$), and they can be transformed away by a continuous non--differentiable ($C^0$) coordinate transformation \cite{kras2,WhatIsAShx,Nolan:2003wp}. 

Shell crossings singularities can be avoided throughout the evolution of dust layers by suitable choices of the initial data. In Szekeres models the necessary and sufficient condition to prevent shell crossings can be simply stated as,
\begin{equation}\Gamma -\bW > 0\quad \hbox{for all}\,\,(t,\vec{r})\,\,\hbox{such that}\quad a>0.\label{noshxtext} \end{equation}
For the case $\Lambda=0$ condition \eqref{noshxtext} can be expressed in terms of the initial functions, but for the general case with $\Lambda>0$ the fulfilment of this condition must be verified numerically (see Appendix \ref{AvoidanceofShx}).

The emergence of shell crossings ({\it i.e.} caustics) mark the onset of virialisation processes (phase mixing and violent relaxation) characteristic of collisionless systems (whether cold dark matter WIMP's or baryons), which indicates the breaking down of a dust continuum as an idealised matter--energy model ~\cite{PadmanabhanBook, binney2011galactic}.  Nevertheless, we emphasize that the dynamical description that we have provided of the formation of pancake--like structures from the Szekeres dust models (connected to the Zeldovich approximation) is appropriate up to the emergence of these caustics.  The proper description of the dynamics of structure formation beyond these caustics lies beyond the present paper and can be obtained (albeit approximately) by numerical N-body simulations, see Sec. 4.10.3 of \cite{binney2011galactic} for details. 
%
%
%The shell crossings indicate the breaking down of dust as an idealised model for matter--energy sources. In turn, shell crossings 
%can be interpreted as the starting point of a highly complicated 
%virialisation process leading to the formation of stable structures, whose proper description is beyond the dynamics of the Szekeres dust 
%models~\cite{PadmanabhanBook} (see also Sec. 4.10.3 of \cite{binney2011galactic} for details on numerical simulation of the relaxation process).
%%%%%%%%%%%%%%%%%%%%%%%%%%%%%%%%%%%%%%%%%%%%%%%%%%%%%%%%%%%%%%%%%%%%%  
In the following we explore the conditions to obtain an evolution free from shell crossings, 
or at least for shell crossings forming sufficiently close to the Big Crunch so that they are covered and hidden away 
by an apparent horizon.     

\subsubsection{Concavity inversions}

The concavity associated with an inhomogeneity (whether density has the local shape of a ``clump'' or ``void'') is closely related with 
the local sign of the ``radial'' coordinate derivatives ($\partial/\partial r$) of the density, which is in turn related with the local sign of 
the density fluctuation $\Drho$. Hence, a local maximum (minimum) in $\rho$ will indicate both a upward (downward) local 
concave profile and an overdensity (underdensity or void).  
Since the density fluctuation can change its sign along the time evolution of the model, local concavity inversions 
(from clumps to voids or vice versa) can occur indicating that local maxima evolve into local minima and vice versa.  
The conditions for these local concavity inversions follow from the existence of solutions of $\DDrho=0$ (or $\Drho=0$): 
%The conditions for these local concavity inversions have been discussed in detail in \cite{Sussman:2010ew} and such inversions are 
%reached when $\DDrho=0$ (or $\Drho=0$): 
%
\ba
&{}&\DDrho=0 \Leftrightarrow \label{InvConConds}
\\
&{}& \! \! \!
H_{qi} \left(\Psi_q-\Psi_{qi}\right)+\frac{2}{3} \left( 1- \frac{H_{qi}}{H_q}\right)=-\frac{\drho_i}{3 \left(\drho_i-\frac{3}{2} \dKK_i\right)},
\nonumber
\ea
with $\Psi_q=H_q (t-\tbb)$. 
As shown in \cite{Sussman:2010ew} for the central extremum in generic LT models, the presence of a decaying mode, or equivalently a 
non--homogeneous Big--Bang time \cite{Sussman:2013qya}, is a required condition for this phenomenon (A statement that is also valid for the evolution of a thick dust shell).
%An important requirement for this phenomenon is the presence of a decaying mode, or equivalently, 
%a non-homogeneous Big-Bang time \cite{Sussman:2013qya}.

\begin{figure}%[H]
\begin{center}
\includegraphics[scale=0.3]{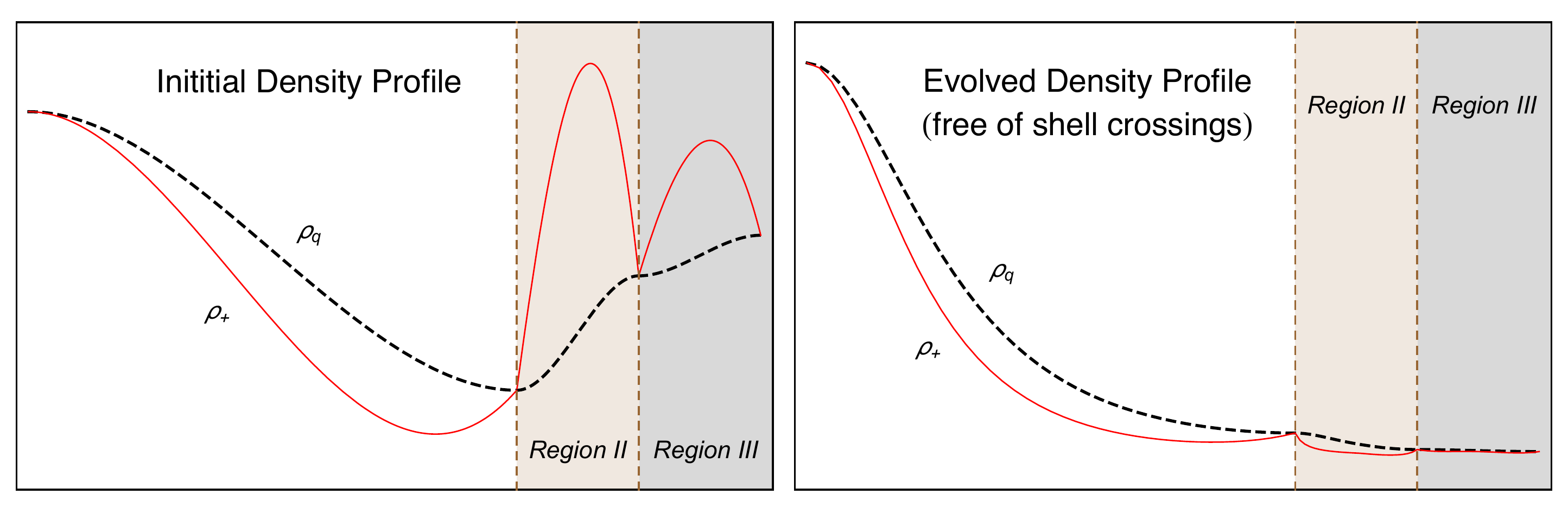}
\caption{
{\footnotesize
{\bf Comparison of density profiles.}  The panels show the profile of the mass density $\rho(r,t,\theta)$ evaluated 
along the curve of the angular maxima of $\bW$ (red and solid)  
and the q--density average $\rho_q(r,t)$ (black and dotted, Eq.~\eqref{Aqdef}). 
The left panel shows a typical initial setup of multiple overdensity structures. 
Evolved with the growing mode, the Szekeres regions II and III will eventually present shell crossings. 
On the other hand, the right panel shows the evolution with a dominant decaying mode which flattens the profile and allows 
for a collapse free of shell crossings (at least at late times).}}
\label{Gprofile}
\end{center}
\end{figure}

In table \ref{tableShX-CI} we examine the possible concavity inversions and shell crossings in the evolution of a dust shell for 
the case $\Lambda=0$.
By looking at all possible combinations of initial conditions, we find that it is impossible to have an evolution of overdensities that keep 
the original concavity profile and collapse onto a BH (a Big Crunch central singularity) without shell crossings at late times\footnote{Shell crossings at very early 
times ({\it i.e.} before $t_i$) are not problematic since they occur out of the range of applicability of the model.}. 
That is, the collapse to a Big Crunch singularity will take place only if the original overdensity is inverted into a void, directly associated to the decaying mode. (See Fig.~\ref{Gprofile} for an illustration of this aspect). 

\begin{table*}[htbp]
\footnotesize
%\centering
\begin{center}
\begin{threeparttable}
\begin{tabular}{|c|c|c|c|c|c|}
\hline
\textbf{Cases}         &\multicolumn{ 2}{|c|}{ \textbf{Description}} & $\textrm{Sign}(\tbb')$     & $\textrm{Sign}(\tcoll')$                        &$\textrm{Sign}(\DDrho_{\textrm{\tiny{coll}}})$ 
\\ 
\hline
I 		               & \multicolumn{ 2}{|c|}{$\dbeta=0\,$ ($\dbeta= -\tiny{\frac{3}{2}} \dKK_0$)}	      & $-\textrm{Sign}(\dKK_0)$  & $-\textrm{Sign}(\dKK_0)$ 		& $\textrm{Sign}(\dKK_0)$ 
 \\ \hline
%
%Case II
II 		& \multicolumn{ 2}{|c|}{$\dKK_0=0\;$ ($\dbeta= \drho_0$)} 		& $\textrm{Sign}(\drho_0)$ 	& $\textrm{Sign}(\drho_0)$ 		& $-\textrm{Sign}(\drho_0)$ 
 \\ \hline
%Case III
%\drho_0-\frac{3}{2} \dKK_0==0
III 	& \multicolumn{ 2}{|c|}{$\dbeta=0$}   & $\textrm{Sign}(\drho_0)$ 		& $\textrm{Sign}(\drho_0)$		 & $-\textrm{Sign}(\drho_0)$ 
 \\ \hline
 %Case IV
 %%
 \multirow{ 2}{*}{IV \tnote{$\dagger$}}  & \multirow{ 2}{*}{$\tbb'<0$}  &       $\dbeta<0$          & $-1$                 & $-1$  		& $\pm1$ 
 \\ 
 & {}    &      $\dbeta>0$          & $-1$                 & $\pm1$  		& $\pm 1$   %\tnote{$\ddagger$}
 \\ \hline
 %
  %Case V
 %%
 \multirow{ 2}{*}{V \tnote{$\dagger$}}  & \multirow{ 2}{*}{$\tbb'=0$}  &       $\dbeta<0$          &  $0$                  &  $-1$		& $+1$ 
 \\ 
                                                             & {}                                        &      $\dbeta>0$          & $0$                    & $+1$  	& $-1$
 \\ \hline
% %
%   %Case VI
% %%
% \multirow{ 2}{*}{VI \tnote{$\dagger$}}  & \multirow{ 2}{*}{$\tbb'>0$}  &       $\dbeta<0$          & $+1$                & $\pm1$ 		& $\mp 1$ 
% \\ 
% %
% & {}    &      $\dbeta>0$          & $+1$                & $+1$  		& $-1$
% \\ \hline

 %
   %Case VI
 %%
 \multirow{ 3}{*}{VI \tnote{$\dagger$}}  & \multirow{ 3}{*}{$\tbb'>0$}  &       \multirow{ 2}{*}{$\dbeta<0$}          & $+1$                & $-1$ 		& $+1$ 
 \\ 
 & {}    &     {}         & $+1$                & $+1$  		& $-1$
  \\
 & {}    &      $\dbeta>0$          & $+1$                & $+1$  		& $-1$
 \\ \hline
\end{tabular}

\begin{tablenotes}
%\item[$\;\;$] In this table $\dbeta\equiv \drho_0-\tiny{\frac{3}{2}} \dKK_0$.

\item $\dbeta \equiv \drho_0-\tiny{\frac{3}{2}} \dKK_0$ and  $^\dagger$ $\drho,\, \dKK_0$ and $\dbeta \neq 0$.

%\item $\dbeta \equiv \drho_0-\tiny{\frac{3}{2}} \dKK_0$.

%\item[$\ddagger$] In this case $\tcoll'>0 \wedge \DDrho_{\textrm{\tiny{coll}}}>0$ are only fulfilled for a very restricted set of initial conditions. Also during almost the whole we have $ \DDrho_{\textrm{\tiny{coll}}}<0$, the inversion of concavity occurs in times very close to the collapse.
%qwerty; \item[2] asdfgh
\end{tablenotes}
\end{threeparttable}
\end{center}
\caption{All possible cases for the evolution of a dust--shell occupying the region $(r^*_1< r < r^*_2)$ in LTB/Szekeres with $\Lambda=0$. 
The table shows the signs of the radial derivative of the big bang and collapse times as well as the sign of the density fluctuation as we approach the big crunch.  
Although positive signs of $\tbb'$ and negative signs of $\tcoll'$ necessarily lead to shell crossings, the shell crossings produced by $\tbb'>0$ occur before $t_i$, out of the range of validity of the model.    }
\label{tableShX-CI}
\end{table*}

\subsection{Apparent horizon}

The apparent horizon is the surface boundary of the region containing  trapped surfaces in which outgoing null geodesic 
congruences present a negative expansion scalar. Although some ``new effects'' appear in quasi--spherical Szekeres models due 
to the lack of symmetry~\cite{kras2,Hellaby2002,BKHC2009,Krasinski:2012hv}, this definition results in the same condition 
as in LTB models: $R=2M$ \cite{Sz75}. Further, in the matching with a Schwarzschild exterior the apparent horizon thus defined 
coincides with the Schwarzschild event horizon\footnote{Note that since the apparent horizon is a quasi--local and foliation--dependent concept, we could have employed another criterium for BH formation instead of the surface $R=2M$, e.g. a concept based on scalar curvature invariants is proposed in~\cite{Coley:2017vxb}.}.

\section{Setting up models of multiple collapsing structures}\label{Sec:SettingUp}

We consider the multiple collapse of cold dark matter structures, with each structure defined by a density maximum of 
quasi--spherical Szekeres or LTB dust models. Such configurations are obtained via a smooth matching 
along the homogeneities spheres of sections of distinct Szekeres spacetimes, and constitute global self--consistent exact solutions of Einstein's equations, as long as the Darmois matching conditions are satisfied along the interfaces of the sections \cite{paperJCAP,israel,taub,bonnor,seno1,kirchner,seno2}.

We look at a specific configuration consisting of a central spherical overdensity described by a section of an LTB spacetime, 
surrounded by two Szekeres shells, each one hosting a non--spherical overdensity, with the most external one smoothly matched to a 
Schwarzschild exterior. 
%(Szekeres spacetimes can be smoothly matched to a Schwarzschild exterior along surfaces of constant 
%$r$~\cite{Hellaby:2007hq}). 
By taking  the dipole parameter $Y=Z=0$ and $X\neq0$, the angular location of these non--spherical overdensities is set at 
$\phi=0$ and $\theta=\pi/2$ (x-axis; see e.g.~Fig.~\ref{DC3d}). 

The initial  density mass is given  in terms of a dimensionless  q--density function ($\mu_{q}$) defined as follows:
\begin{equation}
[\mu_{q}(\chi)]_{i}=\frac{4 \pi [\rho_{q}]_{i}}{3 H_{\ast}^2},\quad\hbox{with}\quad \chi=r/l_s,
\end{equation}
\noindent where $l_s$ and $H_\ast$ are the characteristic length scale and the inverse of the characteristic time scale, respectively. 
In addition, we define the dimensionless curvature ($\kappa_q=\KK_q/H_{\ast}^2$) and rewrite the evolution equations, as well as their analytic solutions, in terms of dimensionless quantities. Proceeding along these lines we have the freedom of choice for both temporal and spatial 
scales. As a consequence a single numerical solution can have various interpretations, 
corresponding to different evolution times ($t-\tbb=(\tau-\taubb)/H_\ast$) and lengthscales ($R(\tau,\chi)=\chi a(\tau,\chi) l_s$). 
%and initial densities. 

Furthermore, we impose initial conditions with a homogeneous Big Bang time ($\tbb^\prime=0$), which sets the initial 
q--curvature through eq. (\ref{tbbmc}).  
This widely used condition is equivalent to avoiding the decaying modes \cite{Sussman:2013qya}.

We find that it is impossible to follow the full evolution of collapsing overdensities without shell crossings emerging before
% they reach 
the Big Crunch. Therefore, we have no alternative but to allow for their presence. 
%of shell crossing singularities. 
In the following we present two possible outcomes from the choice of time and length scales, which result in two different astrophysical 
objects. The initial conditions for these two scenarios are listed in table \ref{tablaIniCondZ7}.
\begin{table}
\footnotesize
\centering
\def\arraystretch{1.0}
%\begin{center}
\begin{tabular}{c|c|c|c|l}

\cline{2-4}
                                        &  $ 0<\chi<\chi_*^1$                 &    $ \chi_*^1<\chi<\chi_*^2$                   &    $ \chi_*^2<\chi<\chi_*^3$                   &  
                                        \\ \cline{1-4}
\multicolumn{1}{|c|}{$\mu_{q i}$}                  &   $\QQ_{1}(\chi)$                &  $\QQ_{2}(\chi)$                     &  $\QQ_{3}(\chi)$                     &  
\\ \cline{1-4}
\multicolumn{1}{|r|}{\multirow{3}{*}{$\bW$}} & \multirow{3}{*}{$X=Y=Z=0$} &    $\kern-4em X=-k_{2}\times$               &         $\kern-4em X=-k_{3}\times$              &  
\\ %\cline{3-4}
\multicolumn{1}{|l|}{}                  &                   & \multicolumn{1}{l|}{$\qquad\times \sin^2 (\frac{\chi-\chi_*^1}{\chi_*^2-\chi_*^1}\pi)$} & \multicolumn{1}{l|}{$\qquad\times\sin^2 (\frac{\chi-\chi_*^2}{\chi_*^3-\chi_*^2}\pi)$} &  %\\ \cline{1-4}
\\ %\cline{3-4}
\multicolumn{1}{|c|}{}                  &                   & \multicolumn{1}{l|}{$Y=Z=0$} & \multicolumn{1}{l|}{$Y=Z=0$} &  \\ \cline{1-4}

\end{tabular}
%\end{center}
\caption{
{\bf{Initial conditions}}. 
The table displays the piecewise definition of the functions 
$\mu_{q i}$ and $\bW$
needed to either integrate the system 
(\ref{FFq1})--(\ref{constraints2}) or evaluate the analytic solutions shown in \ref{AnalyticSols}. 
Functions $\QQ_{i}$ are third order polynomials defined by 
their values 
and vanishing first derivatives at at $\chi_*^0,\chi_*^1,\chi_*^2$ and $\chi_*^3$.
%The normalised coordinates of the matching spheres are $\chi_*^1,\chi_*^2,\chi_*^3=$, with $\chi_*^0=0$. 
 $k_{2}$ and $k_{3}$ are modulling constants of the dipole magnitude.
}

\label{tablaIniCondZ7}
\end{table}

\subsection{An approximate model for a galaxy cluster}
\label{subsection:cluster} 
\begin{figure}%[H]
\begin{center}
\includegraphics[scale=0.4]{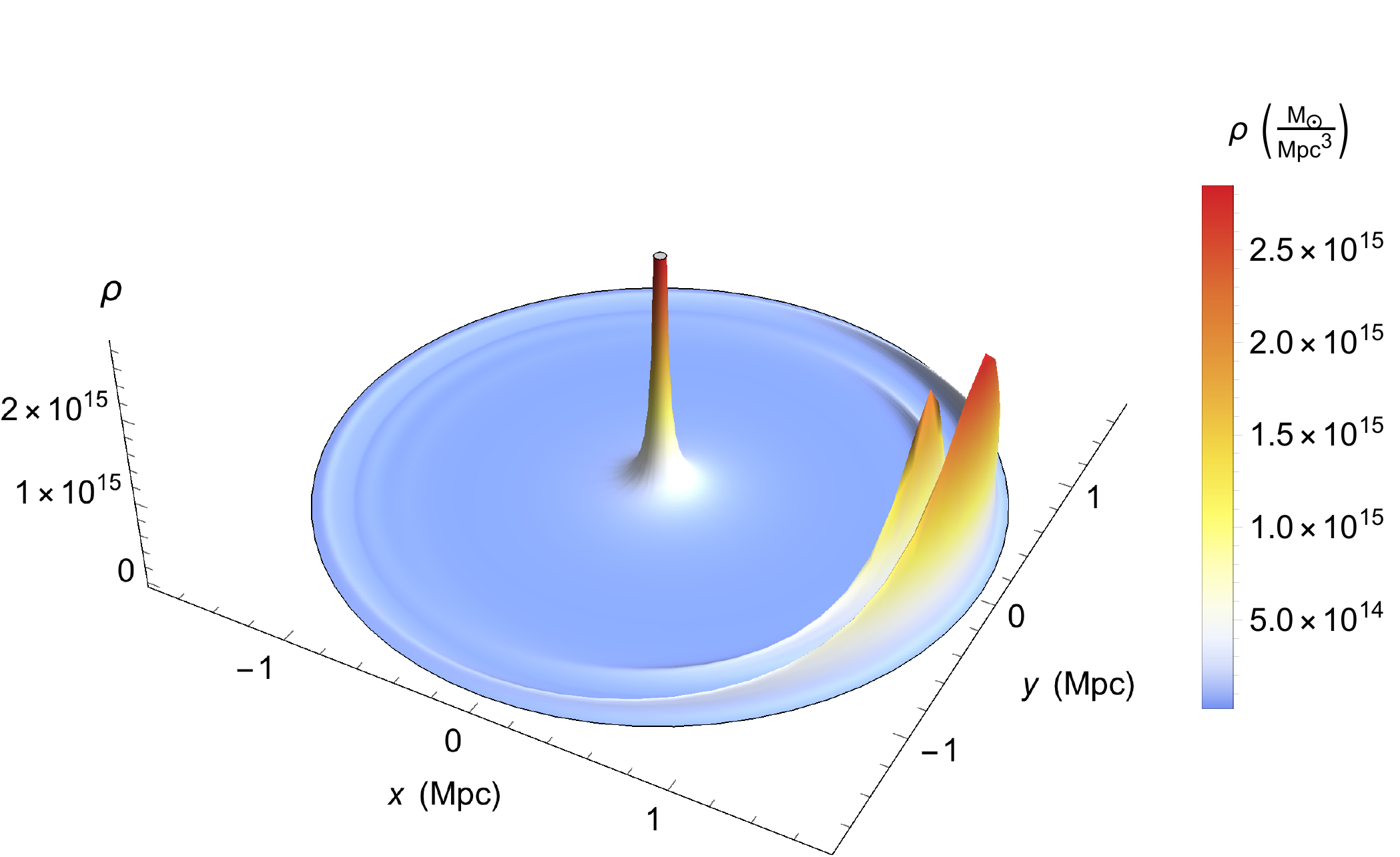}
\caption{
{\footnotesize
{\bf Density mass in the equatorial plane in units of $M_{\odot}/Mpc^3$.} 
Equatorial projection of the density mass distribution at a time close, but before to the time of shell crossings. 
The ``$x$'' and ``$y$'' axes respectively correspond to $R\cos\phi$ and $R\sin\phi$, with $R=a\,r$.
}}
\label{DC3d}
\end{center}
\end{figure}

As a first case, we examine the evolution of a multiple structures configuration from linear conditions 
at a redshift $z=7$
to a present day final configuration of scale $\sim1$ Mpc and mass of $\sim10^{15}$ $M_{\odot}$, which can be compared to a
%present day 
galactic super--cluster\footnote{Following the scheme of table \ref{tablaIniCondZ7}, we set
$\QQ_{1}(0)=1+9.1\times 10^{-3}$, 
$\QQ_{1}(\chi_*^1)=\QQ_{2}(\chi_*^1)=1+10^{-3}$, 
$\QQ_{2}(\chi_*^2)=\QQ_{3}(\chi_*^2)=1+2\times 10^{-3}$ and  
$\QQ_{3}(\chi_*^3)=1+3\times 10^{-3}$ for the piecewise polynomial, $k_1=5.5\times 10^{-1}$ and $k_2=3.4\times 10^{-1}$ for the dipole magnitude and the constant
$H_*=2/(3  t_{z=7})$.}.  
In general terms the evolution proceeds as follows, the structures are initially expanding, then reach (not simultaneously) the turnaround point. Subsequently, at the present 
cosmic time, when the shell crossings first appear, part of the central structure has already collapsed into a hidden spherical singularity. 
Considering the shell crossings as rough estimators of the virialisation time (as interpreted in Sec.~\ref{Sec:Shx}), we can argue that 
the whole set of structures correspond to a cluster that is virialising today and hosting a central BH of 
$\sim10^{9}$ $M_{\odot}$ (see Fig.~\ref{DC3d}).

%
%We take the initial density mass perturbations $\delta\sim10^{-1}$ and specify the initial density and curvature by defining 
%their associated q--scalars as piecewise functions at each interval $\Delta_*^i$, for a sequence of three intervals (see Table \ref{} for more details). The resulting arrangement of structures is matched at the boundary with the Schwarzschild exterior solution.

\begin{figure}%[H]
\begin{center}
\includegraphics[scale=0.4]{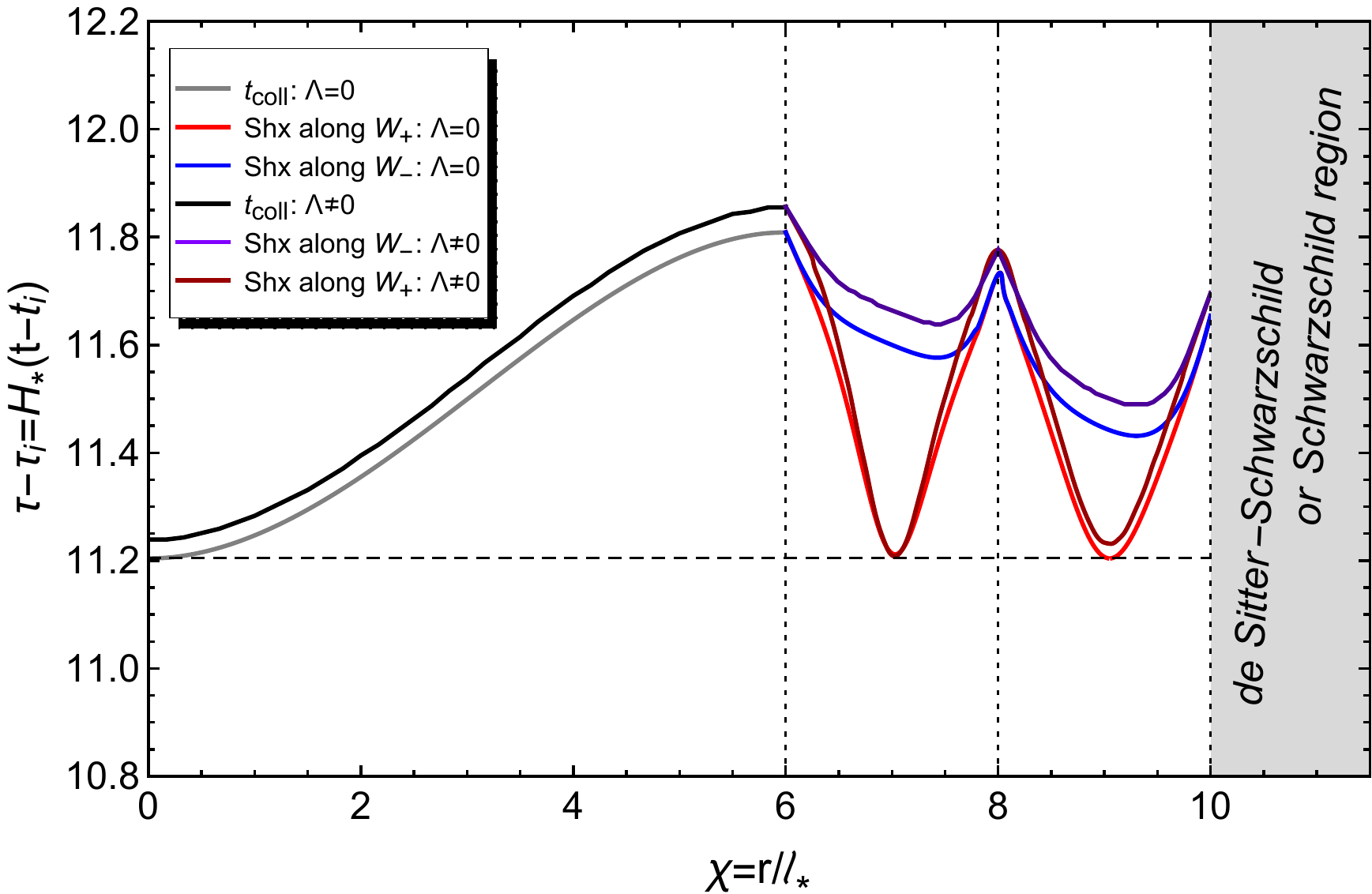}
\caption{
{\footnotesize
{\bf Collapse and shell crossings times.} 
%\stress{
Grey, light red and blue curves respectively represent the times of collapse and shell crossing along the direction of the maxima and minima of the dipole for the case $\Lambda=0$. Black, dark red and violet lines respectively represent the time of collapse and shell crossings considering $\Lambda$.
%  }
}}
\label{CollapseT}
\end{center}
\end{figure}

The times of collapse and shell crossings are shown in Fig.~\ref{CollapseT}. While the collapse time only depends on the comoving radial coordinate, the shell crossing time depends on all the spatial coordinates (cf. Eq.~\eqref{noshxtext}). The red and blue curves represent the shell crossings times along the curves of angular maxima and minima of the dipole function, respectively. We also plot the results of the calculations taking into account the cosmological constant, whose sole effect is to delay the collapse and the shell crossings. 
\subsection{A model for multiple collapse into a black hole}
\label{subsection:multiple}
\begin{figure}%[H]
\begin{center}
\includegraphics[scale=0.36]{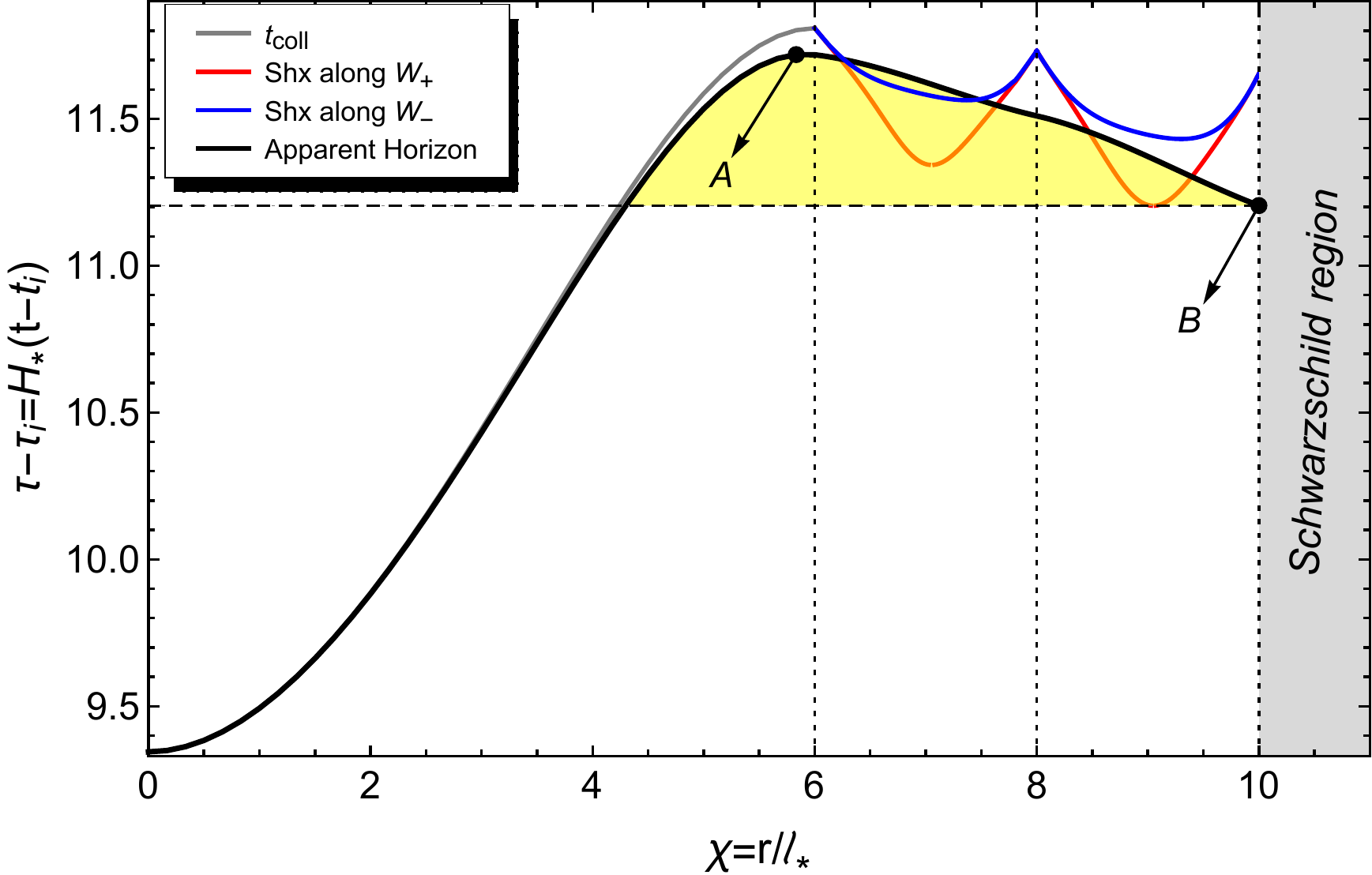}
\caption{
{\footnotesize
{\bf Hiding the shell crossing singularities behind the apparent horizon surface.}
The apparent horizon (black curve) hides the shell crossings thus they are already inside the black hole by the time they appear.
Notice that in the yellow--shaded area the apparent horizon first appears at the point B, where $t_{\hbox{\tiny{AH}}}$ 
has a local minimum, $t_{\hbox{\tiny{AH}}}^{\,(\hbox{\tiny{B}})}$. 
At all times in the interval  ($t_{\hbox{\tiny{AH}}}^{\,(\hbox{\tiny{B}})}$, $t_{\hbox{\tiny{AH}}}^{\,(\hbox{\tiny{A}})}$), the mass
swallowed up by the singularity is 
necessarily smaller than the mass 
%that has disappeared 
into the AH.
}}
\label{gAH}
\end{center}
\end{figure}

\begin{figure*}%[H]
%\includegraphics[scale=0.75]{DC_AIPV.pdf}
%----------------------------------------------------------
\centering
\includegraphics[scale=0.36]{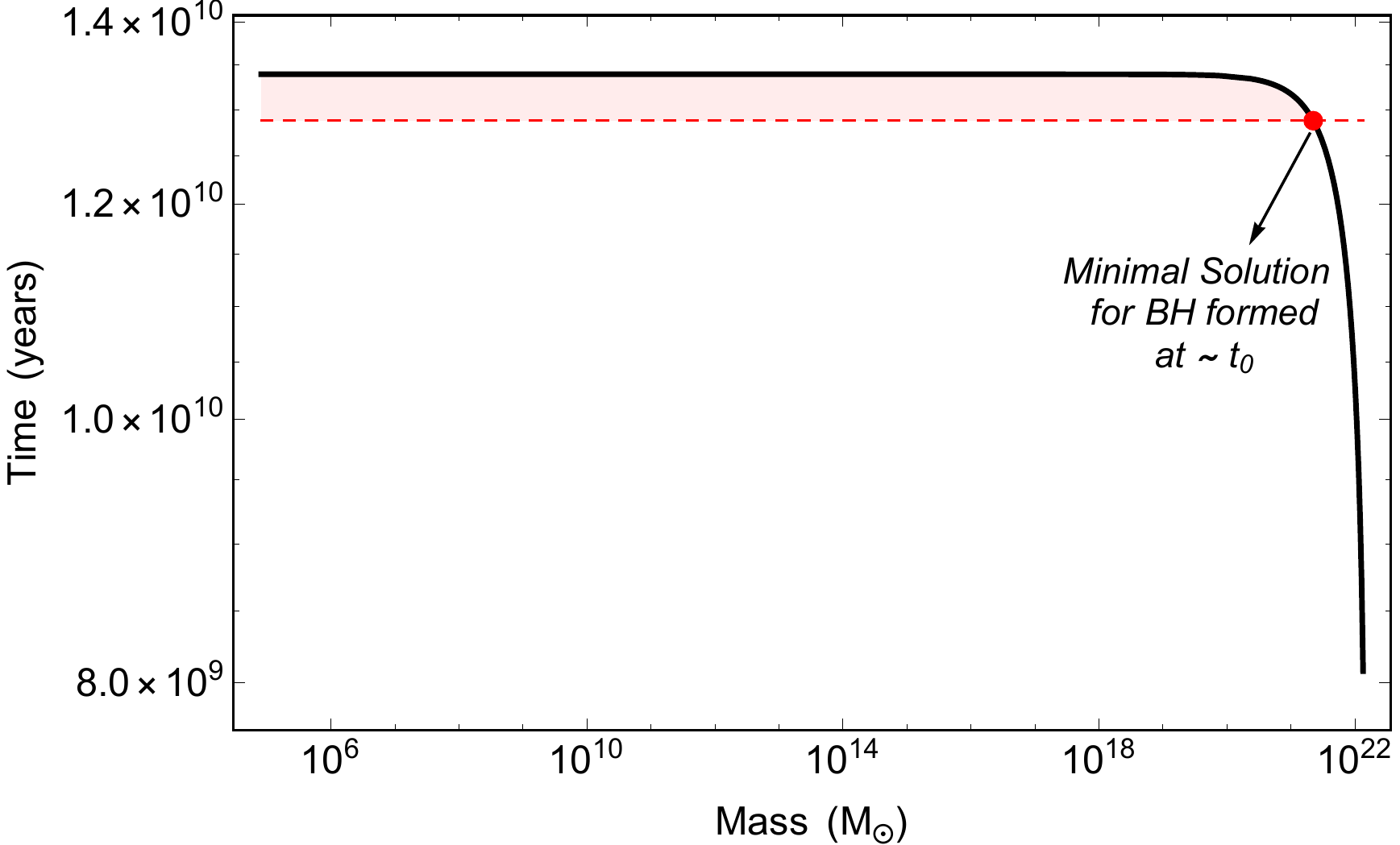}
\quad
\includegraphics[scale=0.34]{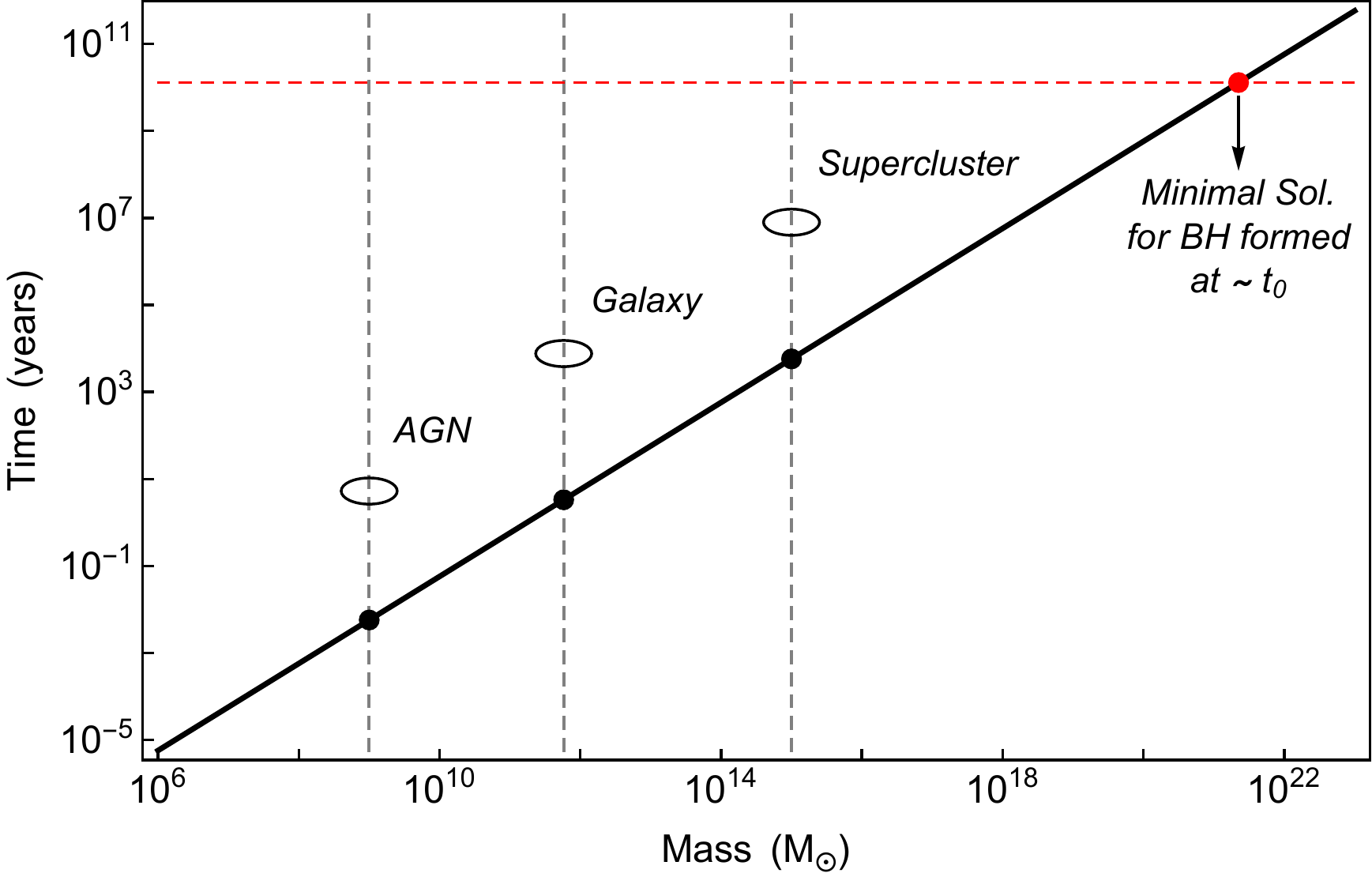}
%__________________________________
\caption{{\bf Black hole formation time as function of the mass.} The left panel shows with a black line the time in which the entire region 
of Szekeres has been hidden behind the Apparent Horizon as a function of black hole mass. For the values of time and masses shaded in 
red, which include the masses of astrophysical black holes, the shell crossings are formed outside the Apparent Horizon. The curve in the 
right panel represents the collapse time as a function the of the mass demanding the condition (\ref{app:horizon}) to be satisfied.
We have also indicated with dashed grey lines the typical masses of Active Galactic Nuclei (AGN, $\sim 10^9 M_\odot$), galaxies ($\sim 10^{11} M_\odot$) and superclusters ($\sim 10^{15} M_\odot$), as a reference. 
}
\label{tvsMass}
\end{figure*}

Another choice is to delay the shell crossings as much as possible ({\it i.e.} as close as possible to the Big Crunch) and accumulate enough mass as to cover them within an apparent horizon surface. 
For that purpose, at the time when shell crossings emerge ($\tau_{\hbox{\tiny{Shx}}}$) the apparent horizon radius must satisfy,
\begin{equation}
\label{app:horizon}
\left[R(\tau,\chi)=2M(\chi)\right]_{(\tau_{\hbox{\tiny{Shx}}}, \chi_*^3)},
\end{equation}
where $R = a\,\chi \,l_s$ is the area distance,
$M=(4\pi/3)\rho_{qi}(\chi \,l_s)^3$ is the quasi--local mass of the whole 
configuration and $\chi_*^3$ marks the boundary between the Szekeres and Schwarzschild regions.
At this surface the apparent horizon coincides with the Schwarzschild event horizon, so by construction, we will have a covered 
singularity\footnote{We took
$\QQ_{1}(0)=1+ 4\times10^{-2} $, 
$\QQ_{1}(\chi_*^1)=\QQ_{2}(\chi_*^1)=1+ 10^{-3}$, 
$\QQ_{2}(\chi_*^2)=\QQ_{3}(\chi_*^2)=1+2\times 10^{-3}$ and  
$\QQ_{3}(\chi_*^3)=1+3\times 10^{-3}$, $k_1=4\times 10^{-1}$, $k_2=3.4\times 10^{-1}$ and the constant
$H_*=2/(3  t_{z=7})$ (see Table~\ref{tablaIniCondZ7}).  }. Fig.~\ref{gAH}
depicts the collapse and shell crossing times, as well as the apparent horizon curve covering the shell crossing singularities.
However, the fulfilment of the condition~\eqref{app:horizon} demands either extremely large values of the overall 
mass or very short collapse times. 
%However, the fulfilment of the condition \eqref{app:horizon} demands extremely large values of the overall mass. 
To illustrate this we consider as a first example  a case with initial conditions at $z=7$  
%and values of $H_\ast$ 
such that the 
shell crossings appeared approximately today.   Then after undertaking an extensive search of initial conditions, 
by trial and error we found that hiding the shell crossing inside the apparent horizon requires large total masses of the order of 
$10^{20} M_{\odot}$.
As shown in the left panel of Fig.~\ref{tvsMass}, for mass values of astrophysical or galactic BHs the shell crossings are formed before being 
covered by the apparent horizon. 
The red--shaded area represents the values of time/mass for which the shell crossings remain uncovered. 
 Conversely one can impose, instead of a fixed shell crossing time, a final mass for the apparent horizon $M= M_{BH}$. 
 Eq.~\eqref{app:horizon} thus fixes the time and length scales for the BH formation. With this choice we find that, just as for the case of 
 PBH formation, the timescales for the collapse are very short (as shown in right panel of Fig.~\ref{tvsMass}). For 
 instance, a typical supermassive BH ($M \approx 10^9 M_\odot$) would collapse 
 in less than a year ($t_{\rm col}- t_{i}\approx 0.005 $ years), while a PBH formed in an early dust-like era at the reheating period is allowed to present a mass of order $M \approx 10^{-16} M_\odot$ \cite{Hidalgo:2017dfp}, and our results show a time of collapse of order $t_{\rm col}- t_{i}\approx 10^{-20} $ seconds. This is in agreement with numerical simulations of PBH formation \cite{Musco:2012au}.

So far we have been concerned with the collapse of multiple pancake--shaped overdensities whose evolution cannot avoid the appearance of shell crossings. The shell crossings, however, can be avoided (at least at later times) by choosing initial conditions whose evolution exhibits concavity inversions which can complete the collapse of dust overdensities towards the central singularity. The resulting mass distribution exhibits a collapse that is qualitatively analogous to that of collapsing LTB models (see Fig.~\ref{Gprofile}) but is sourced by configurations that evolve away of the pure growing mode. 

%On the other hand, when we choose a large central density in order to hide 
%the shell crossings behind an apparent horizon (as illustrated 
%in figure~\ref{gAH}), the evolution resembles that of a multiple structures collapsing onto a single black hole.

\section{Discussion and Final remarks}\label{Sec:DiscFinalRemarks}

In this paper we have studied the collapse of non--spherical structures, modelled by multiple concentrations of pressureless matter. 
We have examined an interesting type of collapse through the joint evolution of a central spherical overdensity 
and neighbour non--spherical structures. 
Specifically, we have looked at the  formation of a spherical apparent horizon, and characterised the possible 
shell crossings that prevent BH formation. 
The latter are interpreted as the breakdown of the dust model and the onset of an intricate 
virialisation process beyond the Szekeres description \cite{PadmanabhanBook}.

%We have presented the full non--linear evolution of inhomogeneities within the Szekeres description. 
%A recurrent phenomenon preventing full collapse of structures is the appearance of shell crossing singularities. 
%As we have shown, the ``central'' region around $r=0$ of quasi--spherical Szekeres models is almost spherically symmetric, 
%and thus is qualitatively equivalent to a LTB dust configuration. 

We have found that the conditions to prevent shell crossings are much more stringent in regions that deviate significantly from homogeneity (cf.~condition \eqref{noshxtext}). Specifically, we have shown that shell crossings cannot be avoided in the collapse of regions where high density pancake--like inhomogeneities evolve in the pure growing mode. 
One way of interpreting the evolution of these multiple configurations is to consider the shell crossings as the onset of virialisation. Note that this characteristic is not exclusive of the non--spherical collapse, since the reported general conditions for the formation of shell crossings (in table \ref{tableShX-CI}) hold even for the case where the dipole is null (LTB case). 

In Sec.~\ref{subsection:cluster}, we exemplify the evolution of a galaxy cluster which starts at $z = 7$ up until the present cosmic time. We simulate the formation of both a central back hole of mass $M \sim 10^{9} M_{\odot}$, and a couple of overdensities which evolve up to the shell crossing time (interpreted as the start of virialisation of galaxy components). 

To examine a full collapse of multiple configurations we have presented examples where the initial conditions delay the emergence of 
shell crossings, so that the latter are covered by an apparent horizon, and remain undetectable to observers in the exterior Schwarzschild 
spacetime. 
In this case, the fact that some dust layers terminate at a shell crossings is practically indistinguishable from the ``real'' collapse 
in which they terminate at the central singularity. 
%This is because, for a given radius, the shell crossings appear at times very close to the 
%appearance of the apparent horizon for the same dust shell. 

The fine--tuned initial conditions needed to build such configuration,  impose constraints on the BH mass and collapse time of the whole structure which are incompatible with astrophysical scales. For example, considering masses of the size of a galactic system we find that extremely large density concentrations are required to obtain a final 
single BH in an astrophysical and/or cosmological time scales, see Fig.~\ref{gAH}. If we assume an initial time around $z\sim 7$, as in our example, the initial distribution of inhomogeneities of galactic mass collapses completely to form a BH of the order of $10^{20} M_\odot$, which is at least five orders of magnitude larger than the typical supercluster mass.  

Alternatively, if we wish to impose a smaller mass for the BH to coincide with (say) a large massive BH in the centre of a 
galaxy ($M\sim 10^9 M_\odot$), then the collapsing timescales become extremely small  ($t_{\rm col}\sim 0.005$ years, see 
right panel of  Fig.~\ref{tvsMass}).  As a consequence, the collapse of non--spherical dust configurations is not an appropriate mechanism to form BHs of astrophysical interest (stellar size or massive BHs in galactic centres or AGNs), not even as a rough toy model level. 
%In contrast, toy models of astrophysical BHs can be devised with LTB models \cite{kras1,kras2,BKHC2009}. 

It is not surprising that self-consistent astrophysical BHs formed from the collapse of non-spherical pancake structures is prevented by shell crossings, as these BHs form from rotating baryonic sources in which hydrodynamical processes become dominant in the regime near the collapse. 

On the other hand, BH formation in the type of Szekeres configurations we are considering is consistent with 
PBHs formation scenarios that involve much smaller masses and very fast collapsing times.  For example, a PBH formed in an early dust-like era of mass $M \approx 10^{-16} M_\odot$, would collapse in $\sim 10^{-20}$ seconds. This is perfectly consistent with PBH formation timescales and our result may complement previous work assessing the formation of PBHs in an early dust-like era \cite{Jedamzik:2010dq,Torres-Lomas:2014bua,Hidalgo:2017dfp,Carr:2017edp}. Our result also argues in favour of recent work on the formation of PBHs from non-spherical configurations \cite{Harada:2015ewt,Harada:2016mhb}.

Finally we comment on the flexibility of the featured model. Our results present enough freedom as to set the mass as an initial condition and preserve it throughout the evolution. Additionally we can manipulate the parameters to set shell crossing times for non--spherical overdensities. This freedom allows us to model either a multiple structures collapse 
(as in the last case studied in Subsection~\ref{subsection:multiple}, and on the other hand, 
allowing for the concavity inversion of inhomogeneities, we can follow their evolution without shell crossing 
singularities up to the time when they cross the apparent horizon.  All the freedom of our model is manifest when working with dimensionless quantities. 
%These features and some of the solutions are presented in Appendix \ref{App:C}. 

\subsection{Acknowledgments}
The authors acknowledge support from research grants SEP-CONACYT 239639 and PAPIIT-UNAM IA103616 
\textit{Observables en Cosmolog\'{\i}a Relativista}. I.D.G. also acknowledges Prof. A. Coley for his hospitality and helpful discussions.
\begin{appendix}

\section{Evolution equations for numerical work}\label{SysOfDiffEq}

The models become fully determined by solving numerically the following set of first order autonomous PDEs (which are effectively constrained ODEs): 
\ba  \dot\rho_q &=& -3 \rho_q\,\HH_q,\label{FFq1}\\
 \dot H_q &=& -H_q^2-\frac{4\pi}{3}\rho_q+\frac{8\pi}{3}\Lambda, \label{FFq2}\\
 \dot\Delta^{(\rho)} &=& -3(1+\Drho)\,\DDH\label{FFq3}\\
 \dot {\textrm{\bf{D}}}^{(H)} &=&  \left(-2H_q+3\DDH\right)\DDH-\frac{4\pi}{3}\rho_q\Drho,\label{FFq4}\\
 \dot a &=& a\,H_q, \label{FFq5}\\
 \dot\GG &=& 3\GG\,\DDH,\qquad \GG=\frac{\Gamma-\bW}{1-\bW}, \label{FFq6}\ea
subject to the algebraic constraints:
\ba
H_q^2 &=&\frac{8\pi}{3}\left[\rho_q+\Lambda\right]-\KK_q,\label{constraints1}\\
\frac{3}{2}\DDKK &=& {4\pi}\rho_q\Drho-3H_q\DDH,\label{constraints2}\ea
where the q--scalars $A_q$ and their fluctuations, $\DDa$ and $\Drho$, are defined in Sec. \ref{qscalarFlucDef}.

\section{Analytic solutions for $\Lambda=0$}\label{AnalyticSols}

For elliptic models, $K>0$, the solution of the quadrature (\ref{quadrature1}) is given explicitly as follows (see \cite{sussbol} for more details and solutions for parabolic and hyperbolic models):
\begin{equation}  t-\tbb  = \left\{\begin{array}{c}
\FF_e(\alpha_{q})/\beta_{qi} \quad\hbox{expanding phase}\,\,H_{qi}>0,\\
 \qquad\qquad\qquad\qquad\qquad\qquad\\
\left[2\pi-\FF_e(\alpha_{q})\right]/\beta_{qi} \quad\hbox{collapsing phase}\,\,H_{qi}<0,  
\end{array}\right.\label{ell}\end{equation}
where $\alpha_q=\alpha_{qi}\,a,\,\, \alpha_{qi} =\frac{3}{4\pi} |\KK_{qi}|/\rho_{qi}$, $\beta_{qi}=\frac{3}{4\pi}|\KK_{qi}|^{3/2}/\rho_{qi}$ 
and $\FF_e$ is defined as 
\begin{equation} 
\FF_e = u\mapsto \arccos(1-u)-\sqrt{u}\sqrt{2-u}.\label{Fe}\end{equation}
The Big Bang, maximal expansion and collapsing times are given by
\begin{equation}  
\tbb = t_i-\frac{\FF_e(\alpha_{qi})}{\bar H_\ast\beta_{qi}}, \;
 \tmax =\tbb +\frac{\pi}{ H_\ast\beta_{qi}},\; \tcoll =\tbb +\frac{2\pi}{H_\ast\beta_{qi}}, 
\label{tbbmc} 
\end{equation}
and the expression for the metric function $\Gamma$, obtained from (\ref{ell}), takes the following form 
\begin{equation} 
\Gamma= 1+\drho_i-3\left(\drho_i-\frac32\dKK_i\right)\left[\Psi_q-\frac{2}{3}\right]-H_q\,r\,\tbb',\label{Gamma1}\end{equation} 
with $H_q$, $\Psi_q$, $\drho_i$ and $\dKK_i$ given by
\ba 
H_q &=& \frac{\dot a}{a}=\pm\frac{\sqrt{\frac{4\pi}{3} \rho_{qi}}\sqrt{2-\alpha_q}}{a^{3/2}},
\\
\Psi_q(\alpha_q) &\equiv& H_q(t-\tbb)=\frac{H_q}{\bar H_\ast}\frac{\FF(\alpha_q)}{\beta_{qi}}
\\
\drho_i&=&\Drho_i|_{_{\bW=0}}=\frac{r}{3}\frac{\rho'_{qi}}{\rho_{qi}},
\\
\dKK_i&=&\DKK_i|_{_{\bW=0}}=\frac{r}{3}\frac{\KK'_{qi}}{\KK_{qi}}.
\ea
For the analysis of the existence of shell crossings, it is worthwhile to re--write $\Gamma$ in the following form, valid during the collapsing phase ($H_q<0$),
\ba 
\Gamma = 1+\drho_i&-&3\left(\drho_i-\frac32\dKK_i\right)\left[|H_q (t-\tcoll)|-\frac{2}{3}\right]\nonumber
\\
&{}&\qquad\qquad\qquad\qquad\qquad+|H_q|\,r\,\tcoll'.\label{Gamma2}
\ea

\section{Dimensionless evolution equations and analytic solutions} \label{App:C}
By introducing dimensionless time, $\tau=H_\ast t$, scale, $R_i=r=\chi \,l_s$, and variables,
\begin{equation}
\mu_q=\frac{4\pi}{3}\frac{\rho_q}{H_\ast^2}, \quad \kappa_q=\frac{\KK_q}{H_\ast^2}, \quad h_q=\frac{H_q}{H_\ast^2}, \quad
\lambda=\frac{8 \pi}{3} \Lambda,
\end{equation}
the evolution equations result in the following dimensionless system,
\ba  \dot\mu_q &=& -3 \mu_q\,\HH_q,\label{FFq1}\\
 \dot h_q &=& -h_q^2-\mu_q+\lambda, \label{FFq2}\\
 \dot\Delta^{(\mu)} &=& -3(1+\Delta^{(\mu)})\,{\textrm{\bf{D}}}^{(h)}\label{FFq3}\\
 \dot {\textrm{\bf{D}}}^{(h)} &=&  \left(-2h_q+3{\textrm{\bf{D}}}^{(h)}\right){\textrm{\bf{D}}}^{(h)}-\mu_q\Delta^{(\mu)},\label{FFq4}\\
 \dot a &=& a\,h_q, \label{FFq5}\\
 \dot\GG &=& 3\GG\,{\textrm{\bf{D}}}^{(h)},\qquad \GG=\frac{\Gamma-\bW}{1-\bW}, \label{FFq6}\ea
subject to the constraints:
\ba
h_q^2 &=& 2 \mu_q+\lambda-\kappa_q,\label{constraints1}\\
\frac{1}{2}{\textrm{\bf{D}}}^{(\kappa)} &=& \mu_q\Delta^{(\mu)}-h_q {\textrm{\bf{D}}}^{(h)},\label{constraints2}\ea
where the arbitrary constants $H_\ast$ and $l_s$ set the time and spatial scales, respectively. As above, $\DDa=A-A_q$
denotes the exact fluctuations and  $\Delta^{(\mu)}=(\mu-\mu_q)/\mu_q$.

On the other hand the analytic solution for the case with $\Lambda=0$, eq. (\ref{ell:dimensionless}), can be rewritten in terms of
dimensionless quantities as
\begin{equation} 
\tau-\taubb  = \left\{\begin{array}{c}
\FF_e(\hat{\alpha}_{q})/\hat{\beta}_{qi} \quad\hbox{expanding phase}\,\,h_{qi}>0,\\
 \qquad\qquad\qquad\qquad\qquad\qquad\\
\left[2\pi-\FF_e(\hat{\alpha}_{q})\right]/\hat{\beta}_{qi} \quad\hbox{collapsing phase}\,\,h_{qi}<0,  
\end{array}\right.
\label{ell:dimensionless}
\end{equation}
where $\hat{\alpha}_q=\hat{\alpha}_{qi}\,a,\,\, \hat{\alpha}_{qi} =|\kappa_{qi}|/\mu_{qi}$, $\hat{\beta}_{qi}=|\kappa_{qi}|^{3/2}/\mu_{qi}$  
and $\FF_e$ was  defined above in eq.~(\ref{Fe}).
Further, the dimensionless big bang, maximal expansion and collapsing times are given by
\begin{equation} 
\taubb = \tau_i-\frac{\FF_e(\hat{\alpha}_{qi})}{\hat{\beta}_{qi}}, 
\quad \taumax =\taubb +\frac{\pi}{\hat{ \beta}_{qi}},
\quad \taucoll =\taubb +\frac{2\pi}{\hat{\beta}_{qi}}, 
\label{taubbmc} 
\end{equation}
and the expression for the metric function $\Gamma$ reads
\begin{equation} 
\Gamma= 1+\dmu_i-3\left(\dmu_i-\frac32\dkappa_i\right)\left[\hat{\Psi}_q-\frac{2}{3}\right]-h_q\,\chi\,\taubb',
\label{Gamma1:dimensionless}
\end{equation} 
where
\ba 
h_q &=& \frac{\dot a}{a}=\pm\frac{\sqrt{\mu_{qi}}\sqrt{2-\hat{\alpha}_q}}{a^{3/2}},
\\
 \hat{\Psi}_q(\hat{\alpha}_q) &\equiv& h_q(\tau-\taubb)=h_q\frac{\FF(\hat{\alpha}_q)}{\hat{\beta}_{qi}}
\\
\dmu_i&=&\Delta^{(\mu)}_i|_{_{\bW=0}}=\frac{\chi}{3}\frac{\mu'_{qi}}{\mu_{qi}},
\\
\dkappa_i&=&\Delta^{(\kappa)}_i|_{_{\bW=0}}=\frac{\chi}{3}\frac{\kappa'_{qi}}{\kappa_{qi}}.
\ea

\section{Avoidance of shell crossings}\label{AvoidanceofShx}

The necessary and sufficient condition to avoid shell crossings can be simply stated as
\begin{equation}\Gamma -\bW > 0\quad \hbox{for all}\,\,(t,\vec{r})\,\,\hbox{such that}\quad a>0.\label{noshx} \end{equation}
From this equation we obtain various necessary (but not sufficient) conditions, such as $\Gamma>0$,\,\, $|\bW|<1$,\,\, 
$0<X^2+Y^2+Z^2<1$ and $|X|,\,|Y|,\,|Z|<1$. For the general case $\Lambda>0$ the necessary and sufficient condition (\ref{noshx}) must 
be verified numerically, but for the case $\Lambda=0$ it can be given in terms of initial value functions. For elliptic models (or regions) 
these conditions are summarised as follows \cite{sussbol}
\ba  
1+\drho_i-\bW\geq 0,\quad \tbb'\leq 0,\quad \tcoll'\geq 0,\\
\drho_i-\frac{3}{2}\dKK_i\geq 0\quad \hbox{necessary not sufficient.}
\ea 
Notice that for the study of a collapsing region the condition $\tbb'\leq 0$ can be relaxed, as it would produce shell crossings that 
can be confined to early cosmic times if $r\tbb'$ is much smaller than horizon distances at $t=t_{i}$. 

\end{appendix}       

%\section*{References}                
%\bibliographystyle{plain}
%\bibliography{SDcites}

\providecommand{\noopsort}[1]{}\providecommand{\singleletter}[1]{#1}%

\end{document}